\documentclass[aps,pra,twocolumn,superscriptaddress]{revtex4-1}
\usepackage{amsmath}
\usepackage{amssymb}
\usepackage{graphicx}
\usepackage{xcolor}
\begin{document}
\title{Exploiting Gaussian steering to probe non-Markovianity due to the interaction with a structured environment}
\author{Massimo Frigerio}
\email{Electronic address: massimo.frigerio@unimi.it}
\affiliation{Quantum Technology Lab $\&$ Applied Quantum Mechanics Group, Dipartimento di Fisica ``Aldo Pontremoli'', Universit\`a degli Studi di Milano, I-20133 Milano, Italy}
\affiliation{INFN, Sezione di Milano, I-20133 Milano, Italy}
\author{Samaneh Hesabi}
\email{samane\_hesabi@yahoo.com}
\affiliation{Department of Physics, Faculty of Science, Shahid Chamran University of Ahvaz, Ahvaz, Iran}
\author{Davood Afshar}
\email{da\_afshar@yahoo.com}
\affiliation{Department of Physics, Faculty of Science, Shahid Chamran University of Ahvaz, Ahvaz, Iran}
\affiliation{Center for research on Laser and Plasma, Shahid Chamran University of Ahvaz, Ahvaz, Iran}
\author{Matteo G. A. Paris}\email{matteo.paris@fisica.unimi.it}
\affiliation{Quantum Technology Lab $\&$ Applied Quantum Mechanics Group, Dipartimento di Fisica ``Aldo Pontremoli'', Universit\`a degli Studi di Milano, I-20133 Milano, Italy}
\affiliation{INFN, Sezione di Milano, I-20133 Milano, Italy}
\date{\today}
\begin{abstract}
We put forward a measure based on Gaussian steering to quantify the non-Markovianity of continuous-variable (CV) Gaussian quantum 
channels. We employ the proposed measure to assess and compare the non-Markovianity of a quantum Brownian motion (QBM) channel, originating from the interaction 
with Ohmic and sub-Ohmic environments with spectral densities described by a Lorentz-Drude cutoff, both at high and low temperatures, showing that sub-Ohmic, high temperature environments lead to highly non-Markovian evolution, with cyclic backflows of Gaussian steerability from the environment to the system. Our results add to the understanding of the interplay between quantum correlations and non-Markovianity for CV systems, and could be implemented at the experimental level to quantify non-Markovianity in some physical scenarios. 
\end{abstract}
\maketitle
\section{Introduction}
It is well acknowledged that the unavoidable interaction of quantum systems with their environment gives rise to decoherence and loss of  information \cite{ref1,ref2}. This phenomenon, however, is not necessarily monotonic in time, and it is possible for some systems to recover previously lost information, at least temporarily. The corresponding backflow of information is one of the characteristic features of 
non-Markovianity, and it attracted large attention in recent years \cite{ref3,ref4,ref5} due to its potential applications to fight decoherence 
in quantum information science. Mathematically speaking, the most general way to capture the concept of non-Markovian quantum dynamics is related to the divisibility and semigroup properties of 
the dynamical map \cite{ref8,ref9} and, in turn, to the degree of accuracy used to derive the master equation describing the dynamics of an open quantum system \cite{ref6,ref7}. On the other hand, from the physical point of view, the non-Markovian character of a quantum map mostly reveals itself through the appearance of memory effects in the dynamics of the system, including information backflow from the environment to the system \cite{ref10,ref11}. 

\par
Different methods have been proposed for the detection and quantification of non-Markovianity \cite{ref12,ref13,ref14,ref15}. Two of the most important measures are the so-called BLP and LFS measures, termed after the names of the authors. The first one is based on the trace distance between two suitably chosen states, whereas the latter involves the mutual information between the studied system and an ancillary one. The non-monotonic behavior of these quantities reveals the non-Markovianity of the channel \cite{ref3,ref16,BLPV16} and may also be used to introduce a measure of non-Markovianity.
\par
Another class of techniques developed to witness quantum non-Markovianity relies upon the effects of non-Markovian dynamics on quantum correlations \cite{ref36,ref38,ref20,ade17}. Among these, the notion of quantum steering, also known as EPR steering in continuous-variable quantum information, identifies a form of correlation between two spatially separated parts of a bipartite quantum system, such that one party may exploit local measurements to "steer" the quantum state of the other one \cite{ref17}. An analog form of quantum steering may be defined in the temporal domain, referring to a single system measured at different times \cite{ref18}. Recently, the role 
of steering in various fields of quantum information, including open quantum systems \cite{ref19} has been investigated. In particular, a non-Markovianity measure has been proposed for discrete variable systems, based on the non-monotonic behavior of temporal steering, and its properties have been analyzed 
\cite{ref20,ref21,ref22}. 
\par 
Continuous variable (CV) open quantum systems \cite{vas1,vas2} 
have an important role in quantum protocols. However, due to the difficulty in evaluations, most of the studies on the non-Markovianity of open quantum systems and the corresponding suggested measures are limited to discrete variable systems. Recently, three non-Markovianity measures have been proposed for Gaussian channels based on fidelity, Gaussian interferometric power and divisibility of map \cite{ref23,ref24,ref25}. Nonetheless, the study of non-Markovianity based on steering has not received the same attention. This is somehow surprising for two reasons: on the one hand because quantum steering is considered a strong type of quantum correlation of some practical interest \cite{ref29} and, on the other hand, because it involves computable quantities. In this paper, we aim at filling this gap by proposing a non-Markovianity measure for Gaussian channels that exploits the concept of Gaussian steering. 
\par
We apply the newly proposed measure to address the non-Markovian properties of a relevant Gaussian channel, namely the quantum Brownian motion channel, originating from the interaction with Ohmic and sub-Ohmic environments characterized by a Lorentz-Drude cutoff. In order to assess the properties of the sole channel, we assume that the probe is initially in a quantum correlated, pure bipartite state, i.e. a two-mode squeezed vacuum state. 
\par
The paper is structured as follows. Two-mode Gaussian states are briefly reviewed in Section II, whereas the concepts of quantum steering and Gaussian steerability are summarized in Section III, where we also introduce our non-Markovianity measure in three variants. The Gaussian channel under investigation and the properties of the environment are detailed in Section IV, while our results about the non-Markovianity of the channel are reported in Section V. In Section VI we discuss the main findings and we look closer at the time behavior of Gaussian steerability to assess the strengths and the weaknesses of the suggested non-Markovianity measures. Section VI closes the paper with a summary from a broader perspective and some concluding remarks. 
\section{Gaussian states}
A two-mode CV quantum system is described on the Hilbert space ${\cal H} = {{\cal H}_{1}} \otimes {{\cal H}_{2}}$, where each ${{\cal H}_{k}}$ is the infinite-dimensional Hilbert space of a bosonic harmonic oscillator. Quadrature operators for each mode are denoted 
by ${q_i} = ({\hat a_i} + \hat a_i^\dag )$ and ${\hat p_i} = -i( \hat{a}_{i} - \hat{a}^{\dagger}_{i} )$, where $a_i$ and $\hat a_i^\dag$ are the annihilation and creation operators, respectively, obeying the bosonic commutation relations $[ \hat{a}_j, \hat{a}^\dag_k]=\delta_{jk}$ with $i,j,k= 1,2$. These quadratures can be grouped in the vector of operators $\hat X = \left( {{{\hat q}_1},{{\hat p}_1},{{\hat q}_2},{{\hat p}_2}} \right)$, and
 the  canonical commutation relations may be rewritten as follows \cite{ref24}: 
 \begin{equation}
\left[ {{{\hat X}_k},{{\hat X}_l}} \right] = 2i{\Omega _{kl}},\,\,\,\,\,\,\,\Omega  = \mathop {\mathop  \oplus \limits_{k = 1} }\limits^N \omega \,\,\,,\,\,\,\,\omega  = \left( {\begin{array}{*{20}{c}}
0&1\\
{ - 1}&0
\end{array}} \right)\,,
\label{eq1}
 \end{equation}
where $\omega$ is usually referred to as the one-mode standard symplectic form. Two-mode {\em Gaussian} CV states are those that can be fully characterized by the vector of expectation values of the quadrature operators, $\langle \hat{X} \rangle$, and by second
 statistical moments of them:
 \[ \left[ \sigma \right]_{jk} \ = \ \langle \{ \hat{X}_{j} , \hat{X}_{k} \} \rangle \ - \ \langle \hat{X}_{j} \rangle \langle \hat{X}_{k} \rangle / 2    \]
 where $\{ \cdot , \cdot \}$ is the anticommutator and $\langle \hat{A} \rangle = \mathrm{Tr} \rho \hat{A}$. These moments constitute a  matrix $\sigma$, the covariance matrix (CM), which is usually written in the following block form for two-mode states:
\begin{equation}
\sigma^{AB} = \left( {\begin{array}{*{20}{c}}
{ \mathbf{A} }&{ \mathbf{C} }\\
{ \mathbf{C}^T}&{ \mathbf{B} }
\end{array}} \right)
\label{eq2}
\end{equation}
where $\mathbf{A}$ and $\mathbf{B}$ are the 
covariance matrices corresponding to each mode, and $\mathbf{C}$ is the 
correlation matrix between them. The blocks $\mathbf{A}, \mathbf{B}$ and $\mathbf{C}$ give rise to a covariance matrix $\sigma^{AB}$ of a 
\emph{physical} two-mode Gaussian state if and only if the uncertainty relations are satisfied, i.e.  \cite{ref24}:
\begin{equation}
\sigma^{AB}  + i\,\Omega  \ge 0\,.
\label{eq3}
\end{equation}    
In our study, we consider a pure Gaussian state as input of the channels under investigation. Up to local transformations, this choice corresponds to the so-called two-mode squeezed vacuum or twin-beam state (TWB), with a CM given by $ \mathbf{A} = \mathbf{B} =a \mathbb{I}$ and
$ \mathbf{C} =\hbox{Diag}(c,-c)$. Upon employing the usual parametrization for them \cite{qopt} in terms of the real two-mode squeezing parameter $r$, Eq.(\ref{eq3}) is satisfied by writing $a= \cosh 2r$ and $c=\sinh 2 r$ and the corresponding CM of the two modes at the initial time will be denoted by $\sigma^{AB}_{0}$ from now on.

\section{Witnessing and measuring non-Markovianity with Gaussian steering}   
Quantum steering (often called EPR steering for CV states) denotes a form of nonlocal correlations that allow one party of a bipartite quantum system to influence, or to \emph{steer}, the quantum state of the other party using local measurement \cite{ref26}, recalling that post-measurement communication between the parties has to occur for the second one to be able to actually detect this influence, in compliance with the no-signaling theorem. EPR steering stands in between Bell nonlocality and entanglement: steerable quantum states are a subset of entangled states, but not all of them are Bell nonlocal states. Notwithstanding this difference, there is no local hidden variable model on the steered side that can account for the effect of EPR steering\footnote{This fact also provides an alternative definition of quantum steering.}, thus the steerable states do not admit a classical counterpart. Moreover, EPR steering is an asymmetric correlation, unlike the other two \cite{ref27,ref37}. Due to this asymmetry, steering is useful in quantum information processes whose measurement results are not specified by one of the subsystems \cite{ref28,ref29}. In particular, for bipartite Gaussian states, the final state of the steered party does not depend on the outcome of the Gaussian measurement performed by the steering party, but just on the \emph{type} of measurement that was performed \cite{ref32}. Several methods have been devised to demonstrate its asymmetry, both theoretically and experimentally \cite{ref30,ref31}. In addition, various criteria and measures have been discussed for its recognition and quantification, such as the quantity introduced by Adesso et. al. for CV systems \cite{ref32}. According to them, Gaussian steering of mode $B$ by mode $A$ of a two-mode CV state can be quantified by the following measure, named \emph{Gaussian $A \to B$ steerability}: 
\begin{equation}
\mathcal{S}^{A \to B} =  \mathrm{Max} \left\{ 0,\frac{1}{2} \ln{ \frac{\det \mathbf{A} }{\det \sigma^{AB} }} \right\}.
\label{eq4}
\end{equation}
where, as before, $\mathbf{A}$ is the diagonal block in $\sigma^{AB}$ pertaining to mode $A$. 
By exchanging the labels $A$ and $B$, one readily obtains $\mathcal{S}^{B \to A}$, a measure of Gaussian steering of mode $A$ by measuring mode $B$. 
Consider now a two-mode Gaussian state that is described by a CM $\sigma(0)$ at the initial time. Under Gaussian, Markovian CPTP maps acting independently on each one of the modes, the time derivative of the steering measure is always non-positive:
\begin{equation} 
\mathcal{D}^{\rightarrow}_{\mathcal{E} } (t) \ : = \ \dfrac{d \mathcal{S}^{A \to B}( \sigma(t)) }{dt} \le 0
\end{equation} 
where $\sigma(t)$ is the CM at time $t$ and $\mathcal{E}_{t}$ is a Gaussian, single-mode CPTP quantum evolution under scrutiny, acting on the steering mode $A$. This is a simple consequence of the divisibility of Markovian maps and the known fact that $\mathcal{S}^{A \to B}$ cannot exceed its initial value under generic \emph{local} Gaussian operations on the modes and classical communication \cite{schur16}. Similarly, one can define $\mathcal{D}_{\mathcal{E}}^{\leftarrow}$ when $\mathcal{E}_{t}$ acts on the \emph{steered} mode $B$ and $\mathcal{D}_{\mathcal{E}}^{\leftrightarrow}$ for $\mathcal{E}_{t}$ acting independently on both modes: also these quantities will be non-negative if $\mathcal{E}_{t}$ describes a Markovian quantum evolution.  However, if the quantum evolution $\mathcal{E}_{t}$ is not divisible in time, i.e. it is non-Markovian, the map that evolves the state between any two consecutive time instants need not be a CPTP map, and consequently, $\mathcal{S}^{A \to B}$ can behave non-monotonically during the whole time evolution. Therefore, any positive value of $\mathcal{D}^{\star}_{\mathcal{E}_{t}}$, with $\star = \rightarrow, \leftarrow, \leftrightarrow$, indicates the violation of Markovianity for the channel under study \cite{ref20}. Building on this fact, a measure of non-Markovianity can be defined for Gaussian quantum channels. With this goal in mind, recall that, at fixed energy, the maximally correlated two-mode Gaussian state is a TWB state, with two-mode squeezing parameter $r$ fixed by the total energy. Moreover, assuming a fixed energy is necessary to have a well-posed problem in CV systems, since it can become unbounded otherwise. Thus, taking a TWB state with two-mode squeezing parameter $r$ and CM $\sigma^{AB}_{0}$ as the initial state, our measure of non-Markovianity for a single-mode Gaussian channel $\mathcal{E}$ is defined as follows:
\begin{equation}
{\cal N}^{\star} \left[ \mathcal{E}_{t} \right](r)  \ = \int_{\mathcal{D}_{\mathcal{E}} > 0} D^{\star}_{\mathcal{E} } \left( t;  r \right) dt
\label{eq5}
\end{equation} 
where it is assumed that $\mathcal{E}_{t}$ acts on the steering mode, the steered mode or on both of them independently, depending on $\star = \rightarrow, \leftarrow, \leftrightarrow$. In the case of a Gaussian quantum evolution $\mathcal{E}^{(n)}_{t}$ acting on $n$ modes, one can extend the previous definitions in the following way. Consider $n$ modes labeled by $a_{1},...,a_{n}$ and call them subsystem $A$, and additional $n$ modes labeled by $b_{1},...,b_{n}$ composing subsystem $B$. As initial state, one can take the tensor product of $n$ TWB states, each of which involves mode $a_{i}$ from system $A$ and mode $b_{i}$ from system $B$, all with the same two-mode squeezing parameter $r$. Subsystem $A$ will be the steering party, while subsystem $B$ will be the steered party. Using the general definition of Gaussian steerability in terms of the symplectic eigenvalues of the Schur complement \cite{ref32}, we can define as before $\mathcal{D}_{\mathcal{E}^{(n)}}^{\star}$ and $\mathcal{N}^{\star} [ \mathcal{E}^{(n)}_{t} ] (r)$ for $\star = \rightarrow, \leftarrow, \leftrightarrow$ and $\mathcal{E}^{(n)}_{t}$ acting respectively on subsystem $A$, subsystem $B$, or independently on both of them. This achieves the goal of defining a generic measure of non-Markovianity for any Gaussian quantum evolution. In the following, we will test $\mathcal{N}^{\star}(r)$ by studying its behavior for Gaussian channels arising from interactions with structured environments. \\

\section{The channel}
This section is devoted to a description of the Gaussian channel studied in this work, the \emph{quantum Brownian motion} (QBM) channel.  The CV system of interest is a quantum harmonic oscillator with frequency $\omega_0$  and unit mass, and it is coupled to a bath consisting of an ensemble of harmonic oscillators. \cite{ref33}. 
An initially factorized state for the system and the bath evolves unitarily under the Hamiltonian of the channel, which is \cite{ref33}: 
\begin{align}
\label{eq6}
\hat{H}_{1} \ &= \frac{{{ {\hat{p}}^2}}}{2} + \frac{1}{2}\omega _0^2{ { \hat{q}}^2} + \sum_{n} \left( \frac{\hat{P}_{n}^{2} }{2 m_n} \ + \ \frac{1}{2} m_n \omega _n^2 \hat{Q}_{n}^{2}  \right) \ \ + \\
& + \ \alpha \hat{q} \sum_n {{k_n}{ \hat{Q}_n}} \notag
\end{align}
where $\hat{p}$ and $\hat{q}$ (resp. $\hat{P}$ and $\hat{Q}$) are quadrature operators of the system (resp. of the bath). The relative strengths of interaction and the coupling constant are denoted by $k_n$  and $\alpha$, respectively.
Then the environment is traced out to find the state of the system at each time. The initial state of the bath is assumed to be a thermal state at temperature $T = \beta^{-1}$ with respect to the corresponding free Hamiltonian of the bath. 
Assuming weak coupling and secular approximation, the master equation can then be recast in this form \cite{qopt} \footnote{We work with a choice of units such that $\hbar = k_{B} = 1$}:
\begin{align}
\label{eq7}
\dfrac{{d\rho }}{{dt}} \  = & \  \dfrac{{\Delta (t) + \gamma (t)}}{2}\left[ 2 \hat{a} \rho \hat{a}^\dag - \hat{a}^\dag \hat{a}\rho  - \rho \hat{a}^\dag \hat{a} \right]  \ + \\
 & + \ \dfrac{{\Delta (t) - \gamma (t)}}{2}\,\left[ 2 \hat{a}^\dag \rho \hat{a} - \hat{a} \hat{a}^\dag \rho  - \rho \hat{a} \hat{a}^\dag  \right] \notag
\end{align}
The time-dependent terms $\Delta (t)$ and $\gamma (t)$ represent the diffusion and damping coefficients, respectively. Assuming a thermal environment at temperature $T$, described by a spectral density $J(\omega )$, these coefficients $\gamma (t)$ and $\Delta (t)$ can be computed by \cite{ref1,ref33}: 
\begin{widetext}
\begin{align}
\gamma (t) & = {\alpha ^2}\int_0^t d\tau \int_0^\infty  d\omega J(\omega )\sin (\omega \tau )  \sin ({\omega _0}\tau )
\label{eq15} \\
\Delta (t) & = \alpha ^2  \int_0^t d \tau \int_0^\infty  d\omega \,J(\omega )\coth \left[ \frac{ \omega }{2 T} \right] \cos (\omega \tau ) \cos ({\omega _0}\tau )
\label{eq16}
\end{align}
\end{widetext}
where $\coth \left[ \frac{\beta \omega}{T} \right] = 2 N(\omega) + 1$ and $N(\omega) = ( \exp \left\{ - \beta \omega \right\} )^{-1}$ is the mean number of thermal photons in the bath at frequency $\omega$. Depending upon the form of the spectral density $J(\omega)$ of the environment, these coefficients may or may not describe a Markovian Gaussian map. In particular, it is known that the quantum map described by Eq.(\ref{eq7}) is non-Markovian if $\Delta(t) < \vert \gamma (t) \vert$ at some point in time. In this study, the origin of the purported non-Markovianity is a spectral density for the bath with Lorentz-Drude cutoff \cite{ref1,ref2,ref24}: 
\begin{equation}
\label{eq:spcdens}
J(\omega ) = \frac{{2{\omega ^s}}}{\pi }\frac{{\omega _c^{3 - s}}}{{\omega _c^2 + {\omega ^2}}}
\end{equation}
where $\omega_c$ is the cutoff frequency. In Eq. (\ref{eq:spcdens}), the cases $s=1$ and $s<1$ correspond to an Ohmic and sub-Ohmic spectral density, respectively, while the super-Ohmic case cannot be dealt with Lorentz-Drude cutoff due to ultraviolet divergence of the frequency integrals. Whenever $\omega_{0} / \omega_{c} \ll 1$, at least in the Ohmic case, the dynamics of the system is expected to be Markovian on the ground of the longer characteristic time of the system compared to the relaxation time of the bath. Therefore, in the following we will assume $\omega_{0} > \omega_{c}$ and consider an environment both at low and high temperature, each in the Ohmic $s=1$ and sub-Ohmic $s = \frac12$ scenarios. Explicit formulas for $\gamma(t)$ and $\Delta(t)$ in the Ohmic cases can be derived, while for the $s= \frac12$ sub-Ohmic spectral density we resorted to numerical integration. 
\par
To discuss our measures of non-Markovianity $\mathcal{N}^{\star}$, two situations must be considered; in the first one, only one mode of the initial TWB state exploited to probe non-Markovianity is subjected to the QBM channel, while the second mode undergoes the free unitary evolution. In this first scenario, adopting the notation:
\begin{align}
\label{eqGamma}
 \Gamma (t) \ & = \ \int_0^t 2\gamma (s)ds  \\ 
\label{eqDGamma}
{\Delta}_{\Gamma}  (t) \ & =  \ e^{-\Gamma(t)} \int_{0}^{t} 
e^{\Gamma (s)} \Delta (s) ds 
\end{align}
the evolution of the covariance matrix,  which is the TWB CM $\sigma _0^{AB}$ at the initial time, when mode $A$ is subjected to the QBM channel is as follows \cite{ref24}: 
\begin{align}
\label{eq8}
\sigma _t^{AB}  = & \left[ e^{ - \frac{\Gamma (t)}{2}} \mathbb{I}_A \oplus \mathbb{I}_B  \right]^T   \sigma_{0} ^{AB}\left[ e^{ - \frac{\Gamma (t)}{2}} \mathbb{I}_A  \oplus \mathbb{I}_B \right] \notag \\
&  + \  \Delta_\Gamma (  \mathbb{I}_A \oplus \mathbb{O}_B )
\end{align}
with the obvious changes when the QBM channel acts on mode $B$ instead. 
Finally, to evaluate $\mathcal{N}^{\leftrightarrow}$ both modes are individually and locally subjected to the channel. The evolution of the covariance matrix is now \cite{ref24}: 
\begin{equation}
\begin{array}{l}
\sigma _t^{AB} = \left[ e^{-\frac{\Gamma(t)}{2}}\mathbb{I}_{A} \oplus e^{ -\frac{\Gamma(t)}{2}} \mathbb{I}_{B}  \right]^T    \sigma_{0}^{AB}  \left[ e^{-\frac{\Gamma(t)}{2}} \mathbb{I}_{A} \oplus e^{-\frac{\Gamma(t)}{2}} \mathbb{I}_{B} \right]\\\\
\,\,\,\,\,\,\,\,\,\,\, \,\,\,+ {\Delta}_{\Gamma} ( \mathbb{I}_{A} \oplus \mathbb{I}_{B})
\end{array}
\label{eq9}
\end{equation}
 In the weak coupling regime $\alpha \ll 1$ that was already implicit in the derivation of the master equation, we can expand $\Delta_{\Gamma}$ to first order in $\Gamma (t)$; then, recalling that $\Gamma (t) \sim O(\alpha^2)$ and truncating the expansion up to second order in $\alpha$, we can simply write:
 \begin{equation}
  {\Delta _\Gamma } \simeq \int_0^t {\Delta (s)ds}
 \end{equation}
 
From the structure of Eq.(\ref{eq8}) and Eq.(\ref{eq9}), it is clear that the blocks $\mathbf{A}, \mathbf{B}, \mathbf{C}$ will still be diagonal and we can write them at any time as $\mathbf{A}(t) = a(t)\, \mathbb{I}_{2}$, $\mathbf{B}(t) = 
b(t)\, \mathbb{I}_{2}$, and $\mathbf{C}(t) = c(t)\, \mathrm{diag} (1,-1)$, 
where $\mathbb{I}_{2}$ is the $2 \times 2$ identity matrix. In particular, the reduced CMs for mode $A$ and mode $B$ will stay proportional to the identity at all times, therefore the free evolution of the modes will not have any effect. In particular, the evolved state will be a two-mode squeezed vacuum state (TMSV), whose Gaussian steerability can also be conveniently judged through the recent concept of nonclassical steering \cite{max21}.  Our task thus boils down to computing the three functions $a(t),b(t),c(t)$ for the three scenarios (bath interacting with mode $A$, bath interacting with mode $B$ and independent, identical bath interacting with both modes at the same time) and then the corresponding non-Markovianity measures. Explicit expressions are collected in Appendix A.
\section{Testing the non-Markovianity measures}
Using Eqs.(\ref{eq8},\ref{eq9}) and the definitions in Eq.(\ref{eq5}), we now evaluate the non-Markovianity measures $\mathcal{N}^{\rightarrow}, \mathcal{N}^{\leftarrow}$ and $\mathcal{N}^{\leftrightarrow}$ for Ohmic and sub-Ohmic environments both at low and high temperatures. We assume a cutoff frequency $\omega_{c} = 1$ and we fix $\omega_{0} = 7$ as the frequency of the system modes. Since $\omega_{c}$ is significantly smaller than $\omega_{0}$, we expect to detect a non-Markovian behavior, as previously discussed. The two-mode squeezing parameter of the TWB initial state of the system will be fixed at $r=2$, which is both an experimentally achievable value in quantum optics and a good representative for the whole class of TWB states: indeed we checked that for any $r \gtrsim 2$ the qualitative behavior that we shall discuss is essentially unaltered. 

 \subsection{Non-Markovianity of an Ohmic environment}
\begin{figure}[!ht]
\centering
\includegraphics[width=0.9\columnwidth]{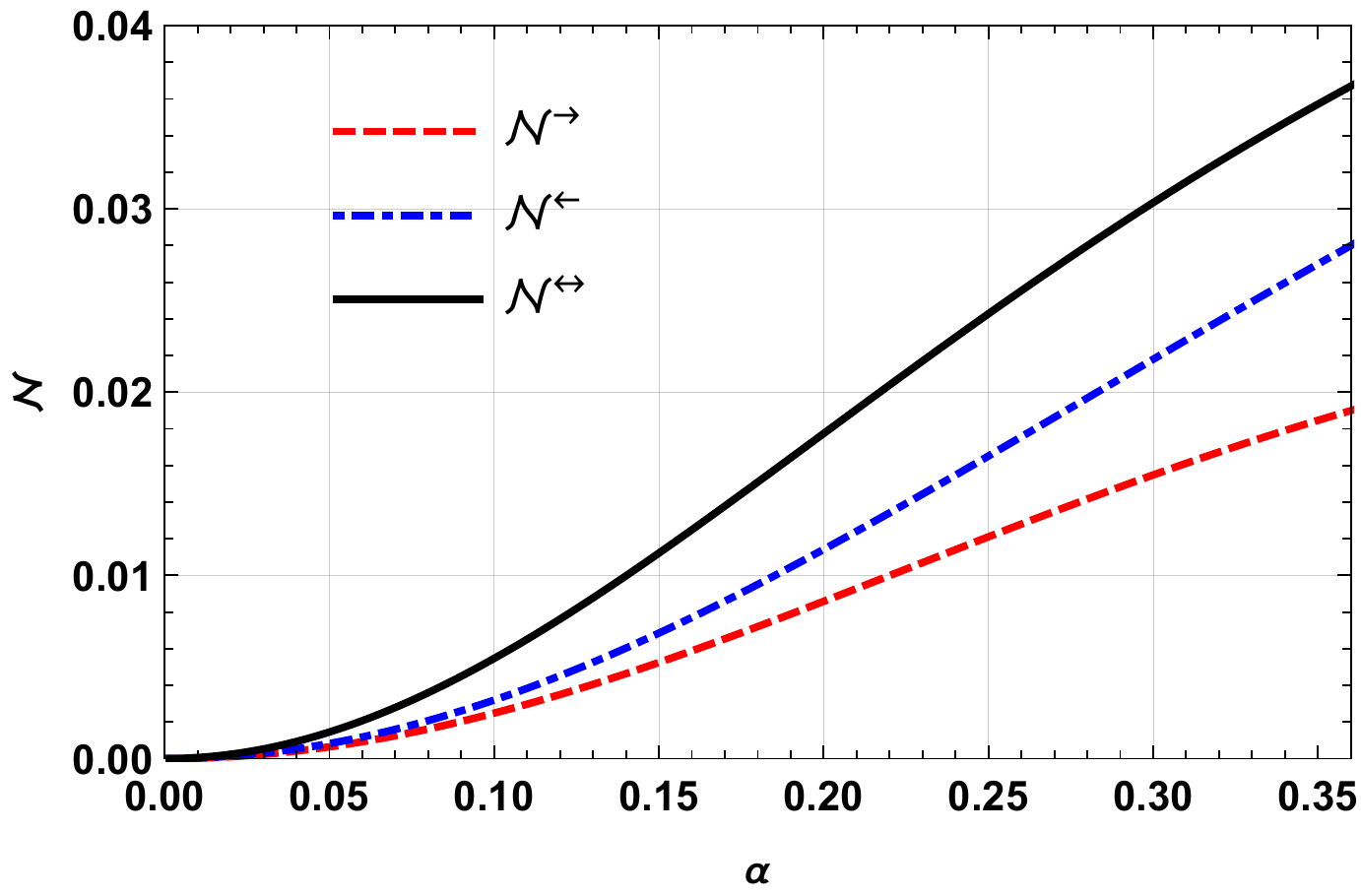}
\caption{\small  Non-Markovianity  (${\cal N}$) vs coupling constant ($\alpha$) for a QBM channel with a structured environment described by an Ohmic spectral density with cutoff frequency ${\omega _c} = 1$ at temperature $T=1.5$. The CV system at initial time is a TWB state with frequency ${\omega _0} = 7$ and two-mode squeezing parameter $r=2$. The three curves correspond to: mode $A$ undergoing the QBM channel evolution and mode $B$ evolving freely ($\mathcal{N}^{\rightarrow}$, red, dashed curve), mode $B$ subjected to the QBM channel and mode $A$ evolving freely ($\mathcal{N}^{\leftarrow}$, blue, dotted-dashed curve) and both modes evolving through two independent, identical QBM channels ($\mathcal{N}^{\leftrightarrow}$, black, solid curve). In all three cases, mode $A$ is the steering mode and mode $B$ is the steered mode. }
\label{fig:1} 
\end{figure}

In this scenario we have $s=1$ in the spectral density, Eq.(\ref{eq:spcdens}). In Fig.\ref{fig:1} the low temperature ($T=1.5$) case is displayed. The measure $\mathcal{N}^{\rightarrow}$ corresponding to the steering mode only (mode $A$) interacting with the environment is the smaller one for all considered values of the coupling constant (red, dashed curve), while $\mathcal{N}^{\leftrightarrow}$ is always the larger one (black, solid curve), corroborating the intuition that the sensitivity of the system to the non-Markovian character of the QBM channel is stronger when both modes can interact with two independent copies of the same structured environment. Moreover, all of the measures are monotonically increasing functions of the coupling and behave as $\sim \alpha^{2}$ in the $\alpha \to 0$ limit. Therefore, for low temperature Ohmic environments, the larger the coupling, the better the estimation of non-Markovianity by Gaussian steering with TWB states. 
\par
\begin{figure}[!ht]
\centering
\includegraphics[width=0.9\columnwidth]{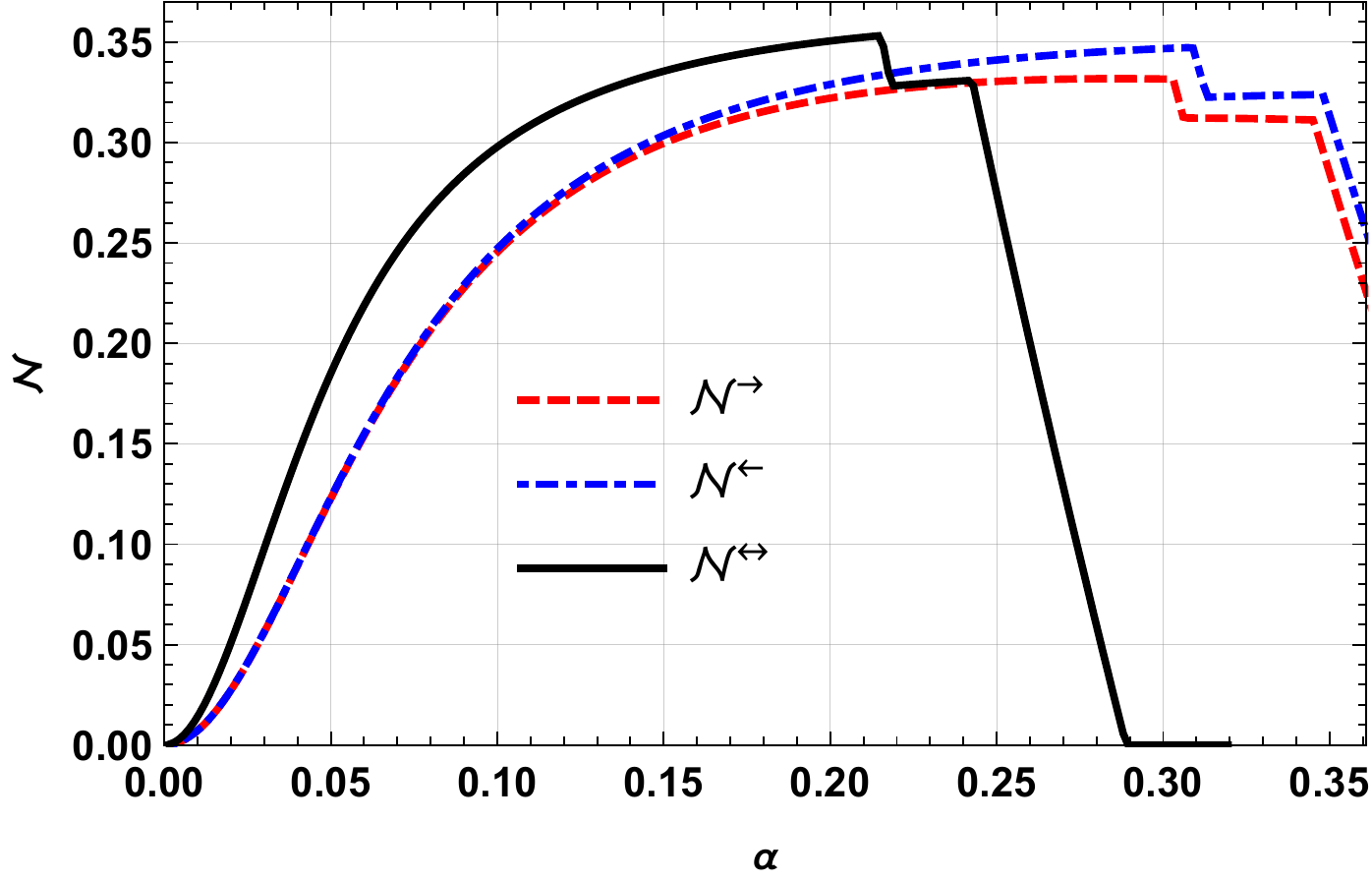}
\caption{\small  Non-Markovianity  (${\cal N}$) vs coupling constant ($\alpha$) for a QBM channel with a structured environment described by an Ohmic spectral density with cutoff frequency ${\omega _c} = 1$ at high temperature $T=100$. The meaning of the curves and the parameters for the system are the same as before. See the 
Discussion Section for a detailed explanation of the sudden jumps.}
\label{fig:2} 
\end{figure}

Considering now a high temperature ($T=100$) Ohmic environment, the same qualitative analysis of the low temperature situation applies for small values of the coupling constant $(\alpha \lesssim 2.1)$, as can be seen in Fig.\ref{fig:2}. The values of $\mathcal{N}^{\star}$ are an order of magnitude higher than those of the low-temperature environment, indicating that not only quantum steering is degraded faster at high temperatures, but it can also be temporarily restored more conspicuously. However, at still higher values of $\alpha$, $\mathcal{N}^{\leftrightarrow}$ suddenly dips to zero with a steep linear slope, and the other two measures soon follow. These are not numerical artifacts, and they will be motivated in the Discussion Section. The upshot is that Gaussian steering can become useless at detecting non-Markovian evolution when there is a strong coupling with a high temperature environment, at least in the Ohmic scenario.  
\subsection{Non-Markovianity of a sub-Ohmic environment}
\begin{figure}[!h]
\centering
\includegraphics[width=0.9\columnwidth]{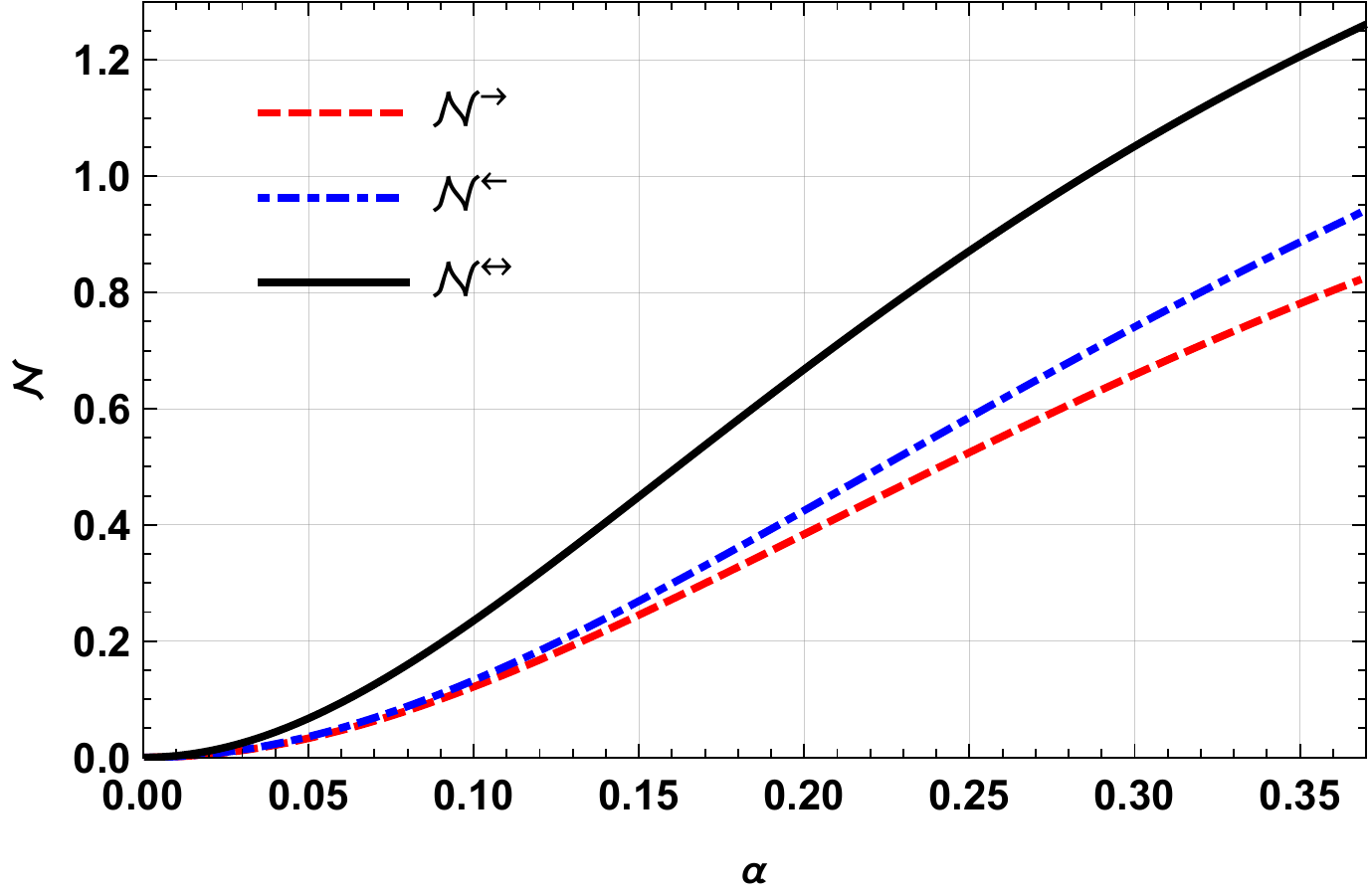}
\caption{\small  The three mon-Markovianity measures  (${\cal N}$) vs coupling constant ($\alpha$) for a QBM channel with a structured environment described by a sub-Ohmic spectral density with $s=\frac12$, cutoff frequency ${\omega _c} = 1$ at low temperature $T=1.5$. The probing CV system is prepared as before. }
\label{fig:3} 
\end{figure}

We now switch to an environment described by a sub-Ohmic spectral density with $s= \frac12$. 
At low temperature ($T=1.5$) the behavior closely mimics the Ohmic case, with $\mathcal{N}^{\rightarrow}$ being the smaller measure and all of them monotonically increasing with $\alpha$ and quadratic in the $\alpha \to 0$ limit, see Fig.\ref{fig:3}. Notice, however, that from the quantitative aspect there is a substantial difference: according to our measures, the non-Markovianity of a sub-Ohmic environment with $s=\frac12$ is about thirty times larger than that of an Ohmic environment at the same temperature and with the same cutoff frequency. 
\par
\begin{figure}[!h]
\centering
\includegraphics[width=0.9\columnwidth]{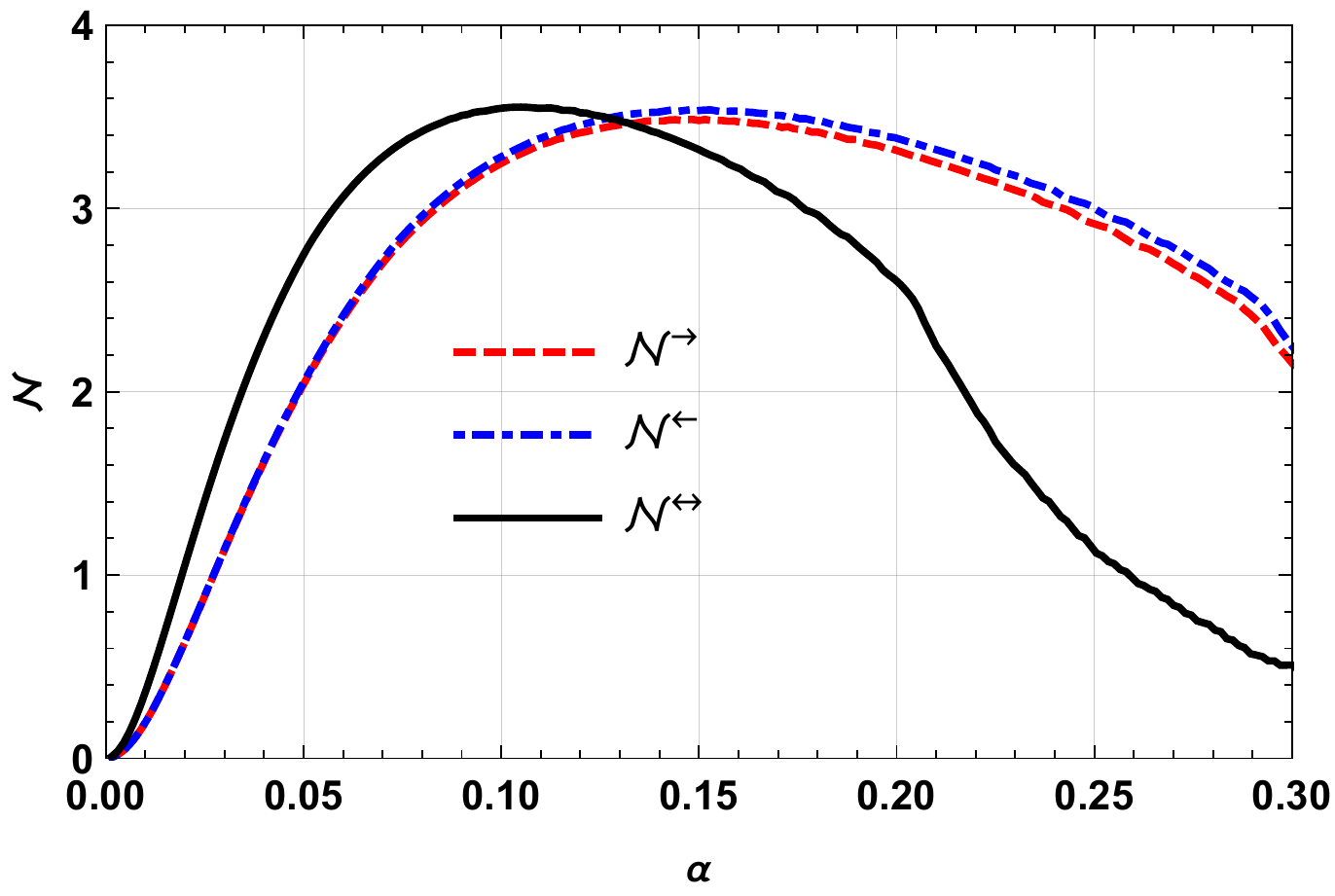}
\caption{\small  The three mon-Markovianity measures  (${\cal N}$) vs coupling constant ($\alpha$) for a QBM channel with a structured environment described by a sub-Ohmic spectral density with $s=\frac12$, cutoff frequency ${\omega _c} = 1$ at high temperature $T=100$. The probing CV system is prepared as before. }
\label{fig:4} 
\end{figure}

Finally, let us consider the QBM channel with a sub-Ohmic environment with $s=\frac12$ and at high temperature ($T=100$). The non-Markovianity measures $\mathcal{N}^\star$ as functions of the coupling constant $\alpha$ are plotted in Fig.\ref{fig:4}. Again the values of all the measures are greater than those for the corresponding low temperature environment, but for $\alpha \gtrsim 0.1$ the plot of $\mathcal{N}^{\leftrightarrow}$ starts to drop gradually, and the other two measures soon follow the same trend. This decrease is considerably slower than that of Fig.\ref{fig:2} for the Ohmic high temperature bath, and an explanation of these trends will be outlined in the Discussion Section. Here too we can conclude that for strong enough couplings with high temperature sub-Ohmic environment, Gaussian steerability is less effective at witnessing the non-Markovian character of this quantum map. 
\section{Discussion}
\begin{figure}[h!]
\centering
\includegraphics[width=0.9\columnwidth]{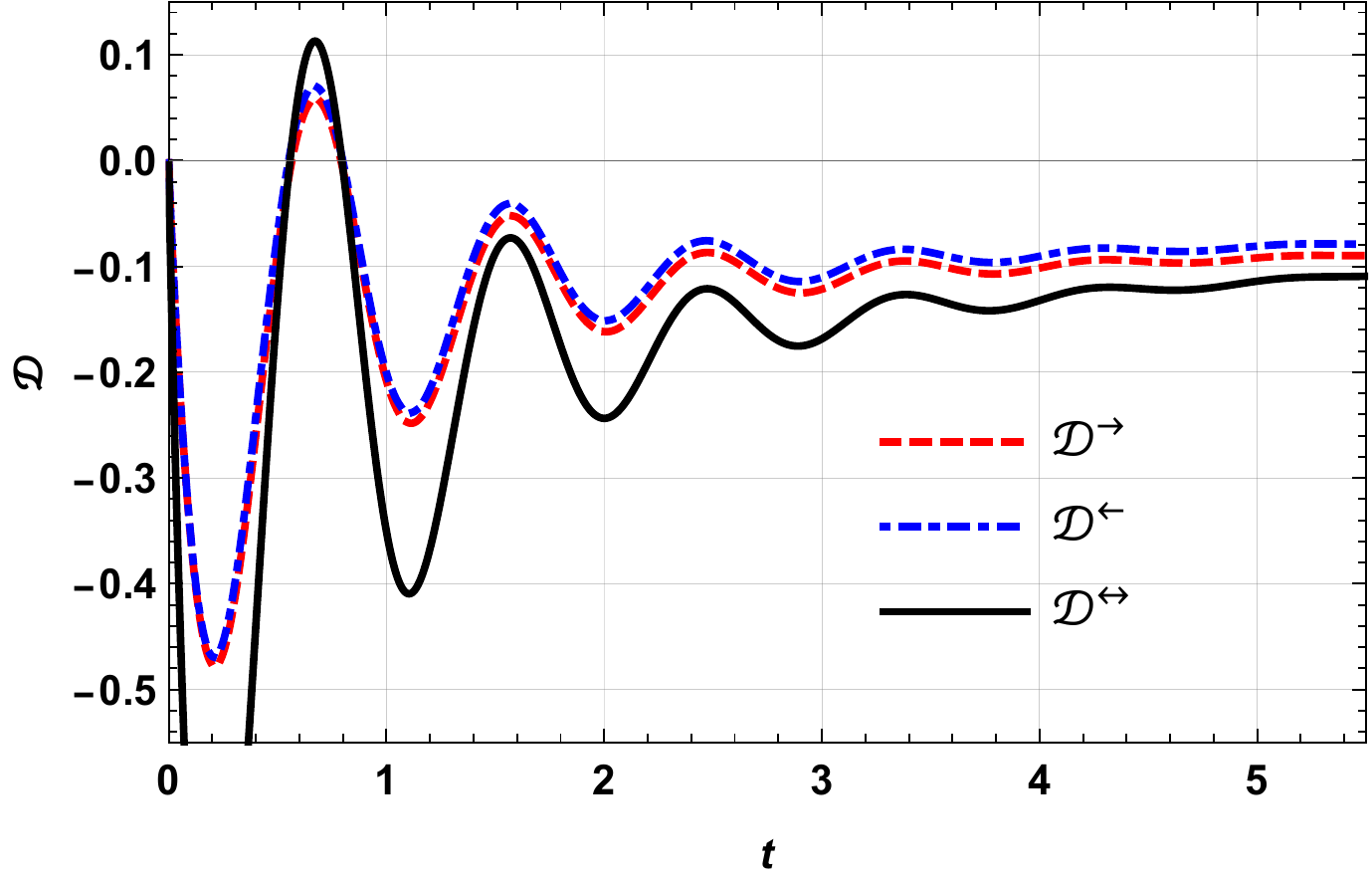}
\caption{\small Time derivative of Gaussian steerability $A \to B$ of a TWB state as a function of time for an Ohmic environment at low temperature ($T=1.5$) and coupling costant $\alpha = 0.15$.   }
\label{fig:6} 
\end{figure}

\begin{figure}[h!]
\centering
\includegraphics[width=0.9\columnwidth]{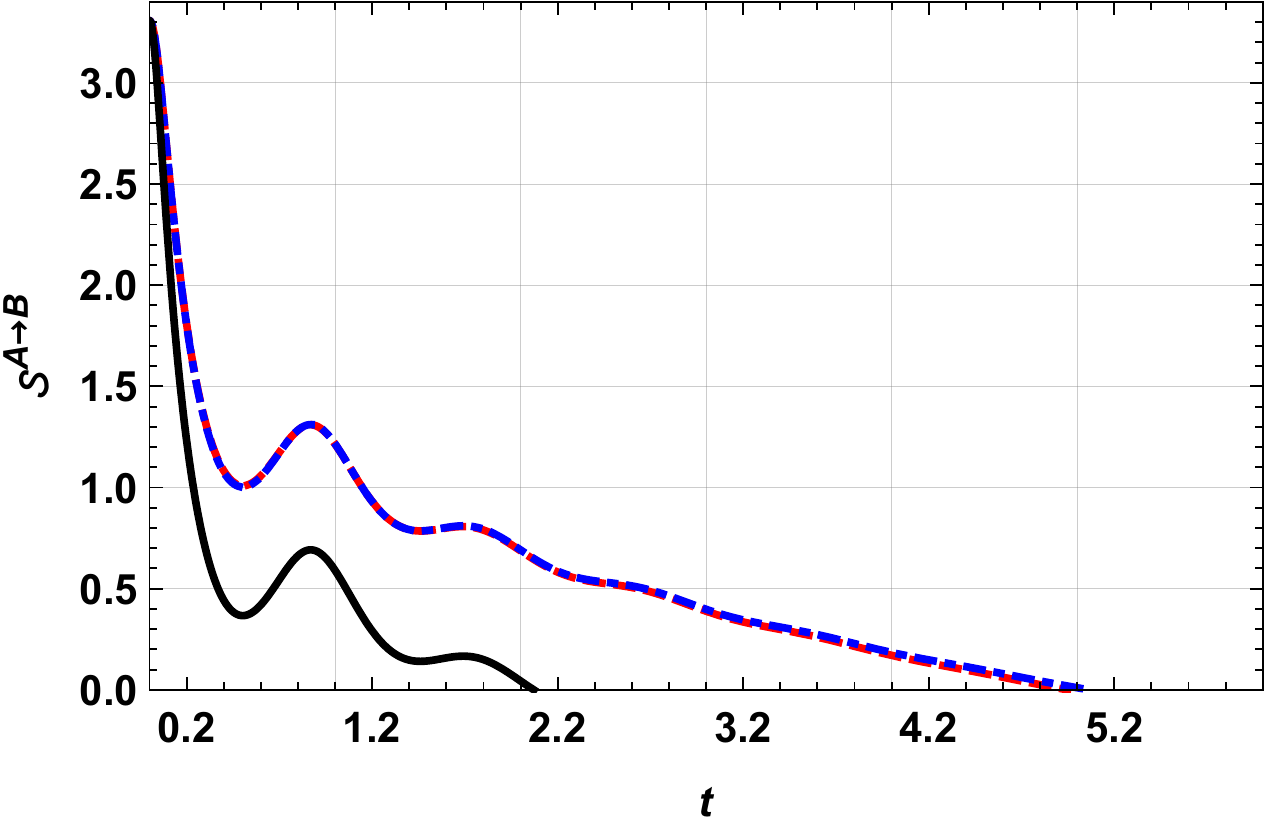}
\caption{\small Gaussian steerability $A \to B$ of a TWB state as a function of time for an Ohmic environment at high temperature ($T=100$) and coupling costant $\alpha = 0.2$.   }
\label{fig:7} 
\end{figure}

As noted in the previous section, the non-Markovianity of the sub-Ohmic environment, according to all of our measures, is considerably higher than that of the Ohmic environment. To have a better understanding of this behavior, and also of the unexpected decrease of $\mathcal{N}^{\star}$ for $\alpha \gtrsim 0.2$ in the case of high temperature environments, it is useful to have a look at the time dependence of the steerability in the two cases.
\begin{figure}[h!]
\centering
\includegraphics[width=0.9\columnwidth]{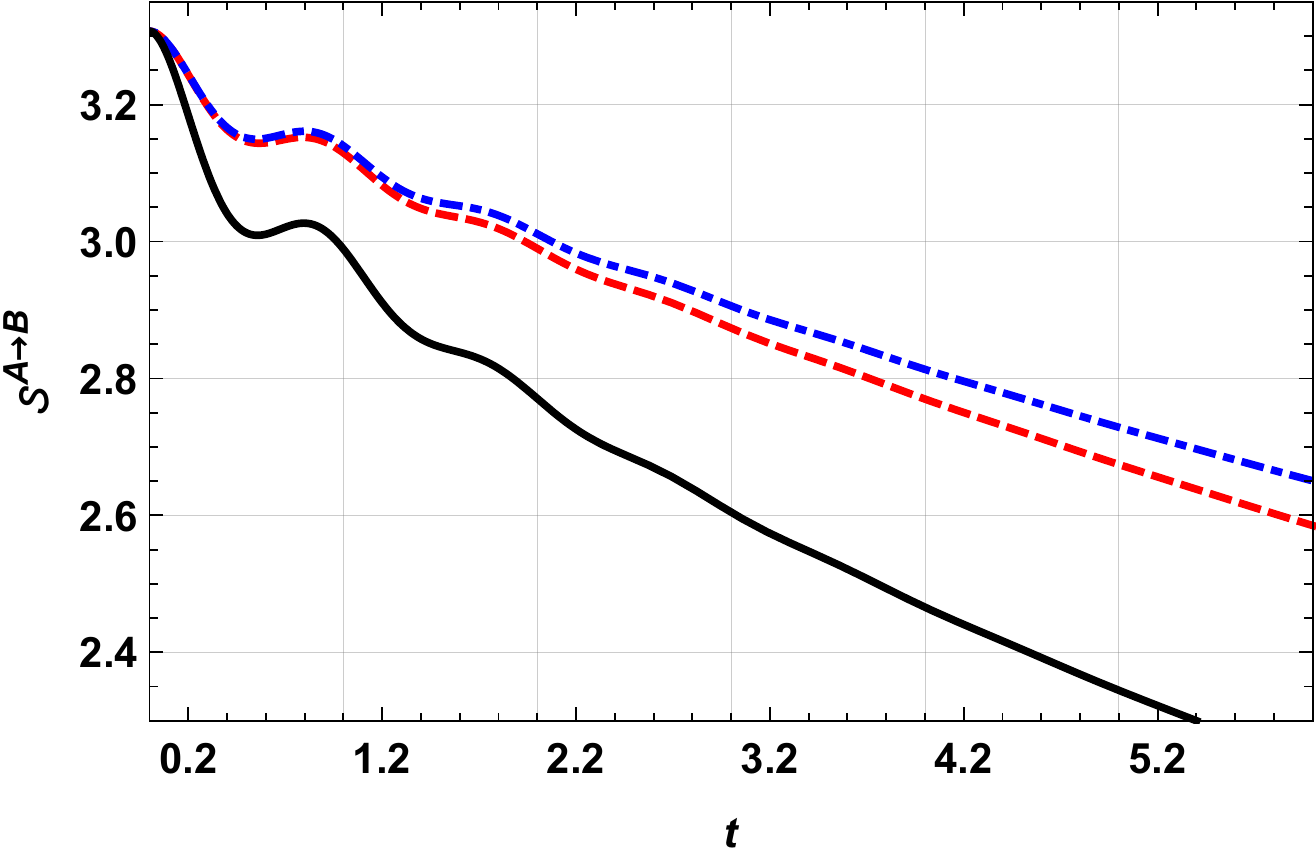}
\caption{\small Time evolution of Gaussian steerability $\mathcal{S}^{A \to B}$ of a TWB state for mode $A$ evolving through a QBM channel with Ohmic environment at low  ($T=1.5$) temperature (red, dashed curve), for mode $B$ subjected to the same QBM channel while mode $A$ evolves freely (blue, dotted-dashed curve) and for both modes undergoing the same channel with two identical, independent environments (black, solid curve). The coupling constant is fixed at $\alpha = 0.2$. Other parameters for the environment and the initial state are again fixed to the previously specified values.  }
\label{fig:5} 
\end{figure}
Fig.\ref{fig:5} displays the Gaussian steerability $\mathcal{S}^{A \to B}$ in the three cases for an Ohmic, low temperature environment ($\omega_{c} = 1$, $T=1.5$, $\omega_{0} = 7$, $r=2$). There is a single time interval, between $t \simeq 0.3$ and $t \simeq 0.8$, during which there is a backflow of steerability (positive slope in the plots). This time interval is approximately the same disregarding of whether the bath interacts with mode $A$, mode $B$, or both modes independently. Moreover, we checked that it is not significantly influenced either by $\alpha$ or by $r$ (at least for $r>2$). The time interval can be better spotted by also plotting the derivatives of the steerability with respect to time, as in Fig.\ref{fig:6}. 
Notice also that the ordering of the curves for the steerability is different from the ordering of the corresponding non-Markovianity measures. Indeed, Gaussian steerability decreases the least when only the steered mode (mode $B$) interacts with the bath (blue, dotted-dashed curve), while it decreases most rapidly when both modes are interacting with independent, identical environments (black, solid curve). 
However, in the latter case, the backflow is also more significant (higher peak in Fig.\ref{fig:6}), yielding greater values of non-Markovianity according to our measures. Indeed, by definition $\mathcal{N}^{\star}$ is just the area between the time axis and the positive arc of the corresponding curve in Fig.\ref{fig:6} (under the red curve for $\mathcal{N}^{\rightarrow}$, the blue curve for $\mathcal{N}^{\leftarrow}$ and the black curve for $\mathcal{N}^{\leftrightarrow}$). 
At higher temperatures, this first time interval is still unchanged, but a second, later interval of increasing $\mathcal{S}^{A \to B}$ appears between $t\simeq 1.4$ and $t\simeq 1.7$ as can be noticed from Fig.\ref{fig:7}, partially explaining the quantitative difference in non-Markovianity between low and high temperature Ohmic environments. As $\alpha$ increases the backflow increases too, but the Gaussian steerability curves of Fig.\ref{fig:7} also slide down towards the time axis and they become zero at earlier times, in compliance with common intuition suggesting that a stronger interaction will degrade quantum correlations faster. 
\par
This explains the trends observed in Fig.\ref{fig:2}: after a certain value of $\alpha$, the Gaussian steerability goes to zero before the end of the second time interval in which backflow can occur (first dip in Fig.\ref{fig:2}). At still higher values of $\alpha$, also the first backflow time interval occurs after the Gaussian steerability has already vanished, and $\mathcal{N}^{\star}$ consequently drops to zero too. This is not seen when the bath is at a lower temperature, because then quantum steering lasts longer and the backflow can always manifest before Gaussian steerability vanishes. 
\par
\begin{figure}[h!]
\centering
\includegraphics[width=0.9\columnwidth]{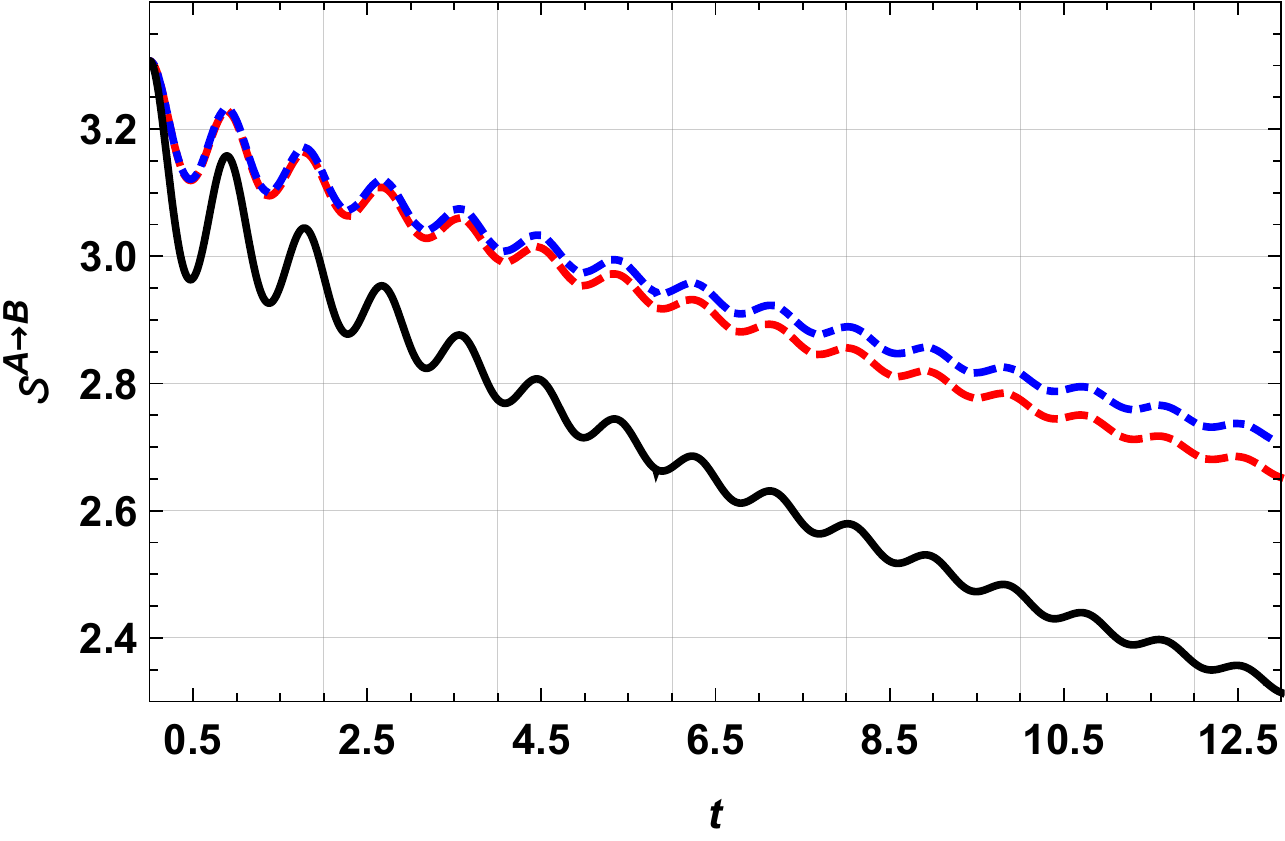}
\caption{\small Gaussian steerability $A \to B$ of a TWB state as a function of time for a sub-Ohmic environment ($s=\frac12$) at low temperature ($T=1.5$) and coupling costant $\alpha = 0.2$.}
\label{fig:8} 
\end{figure}

Fig.\ref{fig:8}, on the other hand, shows that even at low temperature ($T=1.5$), a sub-Ohmic environment with $s=\frac12$ induces cyclic backflows of Gaussian steerability $\mathcal{S}^{A \to B}$ from the bath back into the system, for all three possible ways of coupling the modes of the system with the bath. This gives a quantitative understanding of the difference between non-Markovianity of Ohmic vs. sub-Ohmic environments, and also justifies the slower decrease of $\mathcal{N}^{\star}$ with $\alpha$ at higher temperature observed in Fig.\ref{fig:4}, since now only a minor decrement in non-Markovianity occurs whenever a single time interval of increasing $\mathcal{S}^{A \to B}$ slides past the time when Gaussian steerability vanishes. For the sub-Ohmic case too, this decrease with $\alpha$ is observed just for the high temperature bath. Overall, the issue of decreasing sensitivity to non-Markovianity for stronger couplings with high temperature environments can be overcome by starting with a TWB state having a greater two-mode squeezing parameter $r$, so that Gaussian steerability will not be destroyed before the characteristic time scale of the non-Markovian evolution had time to happen.
\par
As an additional remark, we emphasize that in our analysis we also took into account values of $\alpha$ which could be partially beyond the weak coupling limit that is implicit in the derivation of the Lindblad equation. Therefore it might be the case that in a strict weak coupling limit, only the first portion of the trends discussed so far will be reliable. However, this does not affect the validity of our arguments, because the model described by the final master equation with specified time-dependent diffusion and damping coefficients $\Delta(t)$ and $\gamma(t)$ does not impose any constraint on the value of the coupling constant.
\section{Conclusion} 
In this paper, a general procedure to witness and quantify the non-Markovianity of CV Gaussian quantum maps using Gaussian steering has been introduced. It relies upon the transient restoration of steerability during a non-Markovian time evolution, as probed by twin-beam states with either one of the modes (the steering mode or the steered mode), or both of them, evolving through the channel to be tested. The behavior of these non-Markovianity measures for the quantum Brownian motion channel has been studied for different structured environments characterized by a spectral density with a Lorentz-Drude cutoff, both of Ohmic and sub-Ohmic type and at low and high temperature for each.  
\par
Our results show that, for low temperature environments of both types, non-Markovianity measures increase with the coupling constant. Conversely, at high temperatures, the initially increasing trend is subsided and eventually turned over for sufficiently strong couplings. This can be understood by looking at the time evolution of the Gaussian steerability, which drops to zero too quickly before being able to increase again, when the system is sufficiently coupled to a high temperature environment. From the quantitative point of view, our findings suggest that sub-Ohmic environments lead to appreciably more robust non-Markovianity with respect to Ohmic ones, for the same value of coupling constant, cutoff frequency and temperature. This appears to be caused by a longer, persistent oscillatory time behavior of the Gaussian steerability, which repeatedly experiences short backflows even at longer times in the sub-Ohmic scenario. 
\par
Finally, let us remark that, from an experimental point of view, our work provides a viable opportunity to quantify the non-Markovianity of a Gaussian quantum channel. Indeed, let us consider the two-mode scenario as a simple example, and the measure $\mathcal{N}^{\leftrightarrow}$ as the most sensitive one. The implementation of our proposal requires preparing a TWB state, sending both modes through two independent copies of the channel, and measuring the variances of one quadrature for each mode ($a(t)$ and $b(t)$) as well as the covariance of a quadrature of the first mode with a quadrature of the second mode ($c(t)$), at different times. Using these data, one can compute $\mathcal{S}^{A \to B} (t)$ and consequently $\mathcal{N}^{\leftrightarrow}$, once a reasonably long set of values has been acquired. We hope that our results will foster the understanding of the interplay between non-Markovianity and quantum correlations for continuous-variable systems, adding another facet to the complex and diverse behavior of non-Markovian quantum maps \cite{BLPV16}. 
\begin{acknowledgments}
This work has been supported by Shahid Chamran University of Ahvaz 
via the grant No. SCU.SP98.812.
\end{acknowledgments}

\par
\begin{appendix} 

\section{Time evolution of the entries of the covariance matrix in the three cases} 
In all cases, we have that $\det \mathbf{A}(t) = a^{2}(t)$ and $\det \sigma^{AB}(t) = \left( a(t) b(t) - c^{2} (t) \right)^{2}$, therefore the general formula for Gaussian $A \to B$ steerability becomes:
\begin{equation}
\label{eq:Gsteerappx}
    \mathcal{S}^{A \to B}(t)  = \mathrm{Max} \left\{    0 , \log \left[ \dfrac{a(t)}{a(t) b(t) - c^{2} (t) }  \right] \right\}
\end{equation}

\par
Notice that the non-Markovianity measures $\mathcal{N}^{\star}$ are defined as the integral of the derivative of $\mathcal{S}^{A \to B} (t)$ with respect to time, restricted to the regions where both the derivative and the quantity itself are positive. This means that they can be computed as a sum of differences of the values of $\mathcal{S}^{A \to B}(t)$ at the endpoints of the intervals on which its time derivative is positive, provided that $\mathcal{S}^{A \to B}(t)$ is also positive. The only additional quantities needed are the three functions $a(t),b(t),c(t)$ in the three cases. When mode $A$ interacts with the bath while mode $B$ evolves freely, we find:
\begin{align}
    a^{\rightarrow}(t) \ &= \ e^{ - \Gamma (t) } \cosh (2 r)  \ + \ \Delta_{\Gamma}(t) \\
    b^{\rightarrow}(t) \ &= \  \cosh (2 r)    \\
    c^{\rightarrow}(t) \ &= \ e^{ - \frac12 \Gamma (t) } \cosh (2 r) 
\end{align}
When mode $B$ is subjected to the quantum Brownian motion channel and mode $A$ is left unaffected, instead: 
\begin{align}
    a^{\leftarrow}(t) \ &= \  \cosh (2 r)  \\
    b^{\leftarrow}(t) \ &= \ e^{ - \Gamma (t) } \cosh (2 r)  \ + \ \Delta_{\Gamma}(t)  \\
    c^{\leftarrow}(t) \ &= \ e^{ - \frac12 \Gamma (t) } \cosh (2 r) 
\end{align}
Finally, when both modes independently interact with identical baths, we have:
\begin{align}
    a^{\leftrightarrow}(t) \ &= \  e^{ - \Gamma (t) }\cosh (2 r)  \ + \ \Delta_{\Gamma}(t) \\
    b^{\leftrightarrow}(t) \ &= \ e^{ - \Gamma (t) } \cosh (2 r)  \ + \ \Delta_{\Gamma}(t)  \\
    c^{\leftrightarrow}(t) \ &= \ e^{ - \Gamma (t) } \cosh (2 r) 
\end{align}
Inserting these functions in Eq.(\ref{eq:Gsteerappx}) and computing the finite differences as described before, the non-Markovianity measures $\mathcal{N}^{\star}$ can be evaluated. The functions $\Gamma(t)$ and $\Delta_{\Gamma}(t)$ can be computed from the integrals in Eq.(\ref{eqGamma},\ref{eqDGamma}) once $\gamma(t)$ and $\Delta(t)$ are known. These functions, in turn, can be written explicitly for the Ohmic case:
\begin{widetext}

\begin{align}
    \gamma_{s=1}(t) \ &= \ \frac{\alpha^{2}}{\omega_{0}^{2} + \omega_{c}^{2}} \left[  \omega_{0} -  e^{- \omega_{c} t} \left(  \omega_{0} \cos  \omega_{0} t  + \omega_{c} \sin  \omega_{0} t \right) \right]  \\
    \Delta_{s=1}(t) \ &= \ \frac{\alpha^{2}}{\omega_{0}^{2} + \omega_{c}^{2}}  \left[  \omega_{0}  \coth \left( \frac{\omega_{0}}{2T} \right)  +  e^{- \omega_{c} t} \coth \left( \frac{\omega_{c}}{2T} \right)  \left(  -\omega_{c} \cos  \omega_{0} t  + \omega_{0} \sin  \omega_{0} t \right) \right]  \ + \\
    &  +  \ \alpha^{2} \sum_{n=1}^{+\infty}  e^{-2 n \pi t T } \  \frac{8  n \pi T^{2}  \left( -2  n \pi T \cos  \omega_{0} t   + \omega_{0} \sin \omega_{0} t  \right)   }{( 4 n^2 \pi^2 T^2  + \omega_{0}^{2} )  ( 4 n^2 \pi^2 T^2  - \omega_{c}^{2} )   }      
\end{align}
For all practical purposes, the first few terms of the series in $\Delta(t)$ suffice to very accurately approximate the function. For the $s=\frac12$ sub-Ohmic case, instead, we only computed $\gamma(t)$ in closed form as:
\begin{align}
    \gamma_{s=0.5}(t) \ &= \ \frac{ \alpha^{2} }{2 \sqrt{\omega_{c}} ( \omega_{0}^{2} + \omega_{c}^{2} )} \left[  4 \sqrt{ \omega_{c} \omega_{0}} \mathrm{Cf} \left( \sqrt{  \omega_{0} t} \right) + \sqrt{2} e^{ \omega_{c} t} \mathrm{Erfc} \left( \sqrt{ \omega_{c} t  } \right) \left( \omega_{0} \cos \omega_{0} t  - \omega_{c}  \sin \omega_{0} t \right)          \right. \ + \\
    &  \ \  - \ \left.    \sqrt{2} e^{- \omega_{c} t } \left( 1 + \mathrm{Erfi}   \left( \sqrt{ \omega_{c} t } \right)  \right) \left( \omega_{0} \cos \omega_{0} t  + \omega_{c}  \sin \omega_{0} t \right)         \right]
\end{align}
where $\mathrm{Cf}(z) = \int_{0}^{z} \cos t^{2} \ dt$ is the Fresnel-C function, $\mathrm{Erfi}(z) = -i \mathrm{erf}(i z)$, $\mathrm{Erfc}(z) = 1 - \mathrm{erf} (z)$ and $\mathrm{erf}(z) = \frac{2}{\sqrt{\pi}} \int_{0}^{z} e^{-t^{2}} d t  $ is the error function. Finally, to evaluate $\Delta_{s=0.5}(t)$ we resorted to numerical integration. 
\end{widetext}
\end{appendix}

\bibliography{nmbycvst3}
\end{document}